\documentclass[a4paper]{jpconf}
\usepackage{graphicx}
\usepackage[square,sort&compress]{natbib}
\begin{document}
\title{The nuclear radio structure of X-ray bright AGN\footnote{Based on observations made with MERLIN and EVN.}}

\author{Jens Zuther$^{1,2}$, Sebastian Fischer$^2$, Andreas Eckart$^{2,3}$}

\address{$^1$Max-Planck-Institut f\"ur extraterrestrische Physik, Giessenbachstr., 85748 Garching, Germany}
\address{$^2$I. Physikalisches Institut, Universit\"at zu K\"oln, Z\"ulpicher Str. 77, 50937 K\"oln, Germany}
\address{$^3$Max-Planck-Institut f\"ur Radioastronomie, Auf dem H\"ugel 69, 53121 Bonn, Germany}

\ead{jzuther@mpe.mpg.de}

\begin{abstract}
The physical nature of the X-ray/radio correlation of AGN is still an unsolved question. High angular resolution observations are necessary to disentangle the associated energy dynamics into nuclear and stellar components. We present MERLIN/EVN 18~cm observations of 13 X-raying AGN. The sample consists of Seyfert 1, Narrow Line Seyfert 1, and LINER-like galaxies. We find that for all objects the radio emission is unresolved and that the radio luminosities and brightness temperatures are too high for star formation to play an important role. This indicates that the radio emission in these sources is closely connected to processes that occur in the vicinity of the central massive black hole, also where the X-ray emission is believed to originate in.
\end{abstract}

\section{Introduction}
Among current problems in AGN research are the feedback between the central engine and the host galaxy environment \cite{page_submillimeter_2001}. Related to this is the observation that only about 10\% of the local galaxy population display Seyfert characteristics \cite{ho_search_1997}, which raises questions like: (i) How is the nuclear engine fueled? (ii) What is the relationship between Seyfert activity and circumnuclear star formation? In order to address these questions it is important to disentangle the energy dynamics around the nucleus at high angular resolution. The NIR-to-X-ray spectral energy distributions (SEDs) of Seyfert 1 galaxies are quite flat \cite{2004ASPC..311...37W}. The X-rays are believed to originate close to the accretion disk in a kind of hot corona \cite{1978Natur.272..706S}. Only a small fraction of X-rays can be attributed to a stellar component in AGN \cite{2005ApJ...631..707P}. On the other hand, extended radio emission is most likely produced via Synchrotron processes in radio jets or supernovae \cite{1984ARA&A..22..319B,1992ARA&A..30..575C}. The core radio emission generally exhibits a flat, self-absorbed spectrum, whereas the spectral slope of radio jets is steep. Furthermore, there appears to be a tight correlation between X-ray and radio flux, both for AGN and star-forming galaxies \cite{2007A&A...467..519P}. Such a correlation suggests a common physical nature that links the emission mechanisms of both phenomena. It is, however, not clear what this link is, since both types of radiation are believed to originate on different physical scales (accretion  disk vs. jet). One solution for flat spectrum point sources is possible provided that the emitting region is hot and optically thin \cite{blundell_origin_2007}. In this case, optically thin bremsstrahlung from a slow, dense disk wind can contribute significantly to the observed levels of nuclear emission. In this case, radio and X-ray emission originate from the same regions and the correlation can be attributed to a common disk origin.

The spatial distribution of the radio emission can, therefore, tell us something about the origin and relative importance of the radio emission. Extended but collimated emission is linked to a jet, diffuse emission is usually attributed to star formation, and compact nuclear emission to processes close to the supermassive black hole (SMBH).

\begin{figure}[t!]
\begin{center}
\includegraphics[width=12cm]{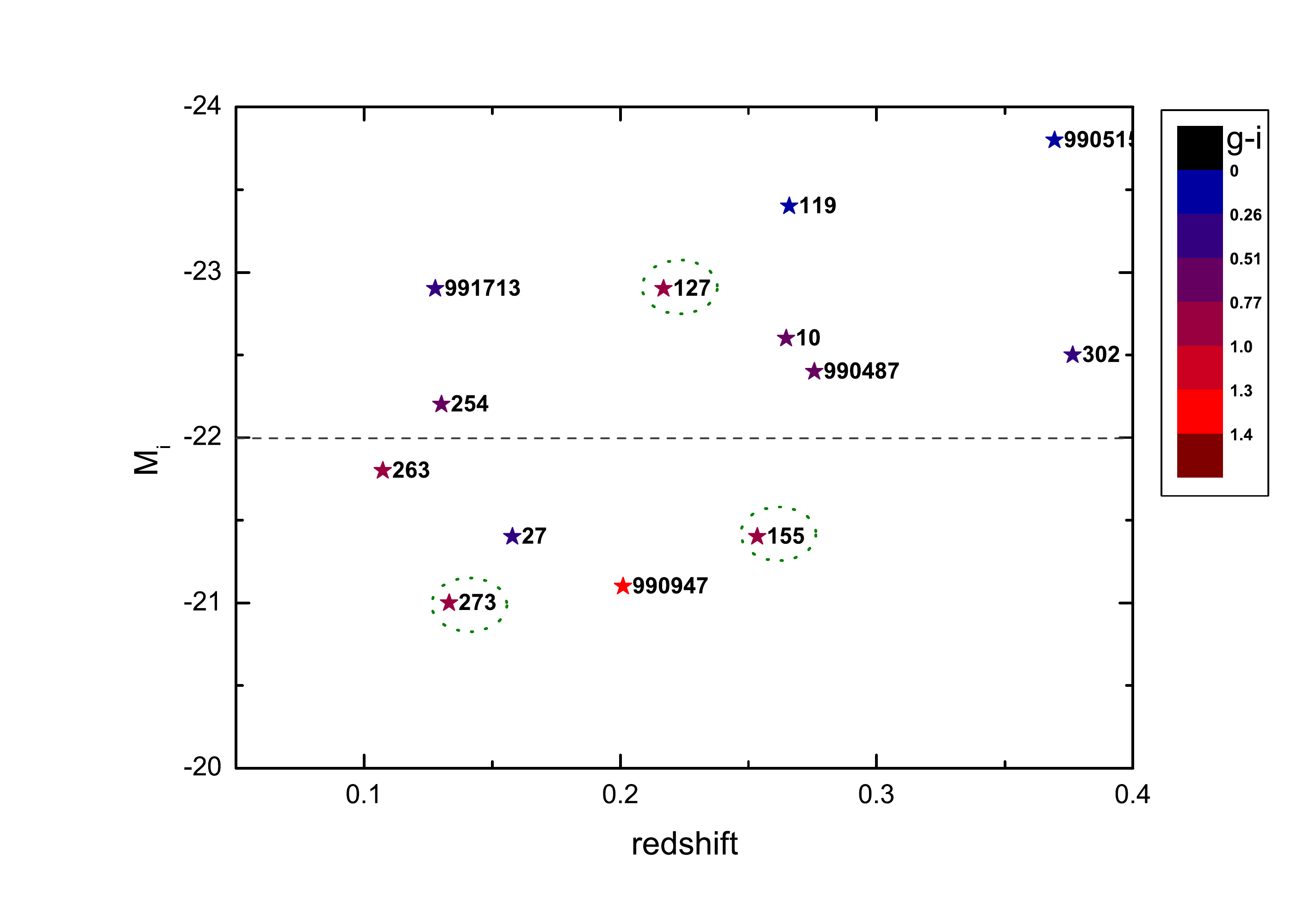}
\end{center}
\caption{Absolute $i$-band PSF magnitude as function of redshift. The dashed, horizontal line marks the demarcation between Seyfert galaxies and QSOs. The color coding of the symbols corresponds to the $(g-i)_\mathrm{PSF}$ color of the targets. The dotted circles indicate sources that are members of galaxy clusters. Numbers are target identifiers \citep[cf][]{zuther08}.}
\label{fig:MagZDist}
\end{figure}

\section{The sample and observations}
In order to tackle the questions posed above, we cross-matched the databases of the Sloan Digital Sky Survey (DR5; \cite{2007ApJS..172..634A}) and the ROSAT All Sky Survey (RASS; \cite{1999A&A...349..389V}). The result is a sample of X-raying AGN that is in part suitable for adaptive optics (AO) assisted follow-up observations in the NIR (i.e. the galaxies have a nearby natural guide star \citep[e.g.,][]{ 2004ASPC..311..325Z,2005sao..conf..375Z,diss07}). Subsequently, we matched the sample with the FIRST database. The sample consists of 13 objects, i.e. 
\begin{itemize}
\item 6 Narrow-Line Seyfert 1 (10, 27, 119, 263, 273, 991713)
\item 3 Seyfert 1 (254, 990487, 990515)
\item 1 Blazar (302)
\item 2 LINER (127, 155)
\item 1 passive (990947)
\end{itemize}
galaxies. The average peak FIRST flux density is about 5~mJy. The redshift and absolute brightness distribution can be seen in Fig. \ref{fig:MagZDist}. All sources, except target 515, are radio quiet or radio intermediate and unresolved on the FIRST scale. Luminosities have been calculated using the concordance cosmology ($h=0.7$, $\Omega=0.3$, and $\Lambda=0.7$).

All targets have been studied with MERLIN at 18~cm. Target IDs starting with 99 have been followed-up with western EVN, again at 18~cm \cite{zuther08}.

\begin{figure}
\includegraphics[width=18pc]{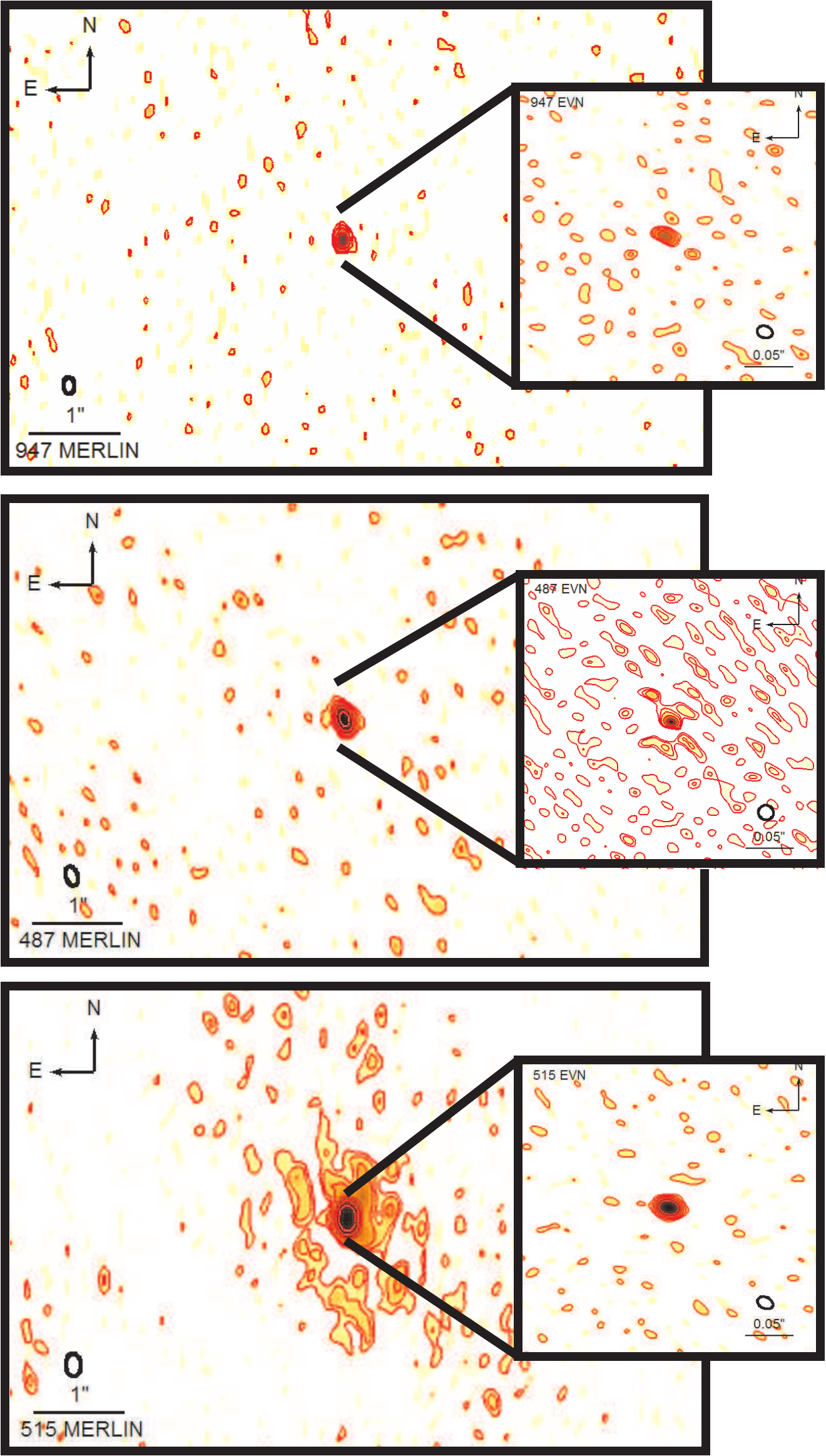}
\begin{minipage}[b]{16pc}\caption{\label{label}18~cm MERLIN maps of the three sources that have also been observed with EVN (insets). The contours shown are multiples of the rms in the images (2, 4, 8, 16 ...). The beam sizes and orientations are indicated by the ellipses. Except source 515, which is a radio loud quasar with a large scale jet, the targets are clearly unresolved.}
\end{minipage}
\label{fig:evnMaps}
\end{figure}

\section{Results}
Both observations were carried out in phase referencing mode. Fig. \ref{fig:evnMaps} presents maps of the objects that have both MERLIN and EVN detections. The average size of the restoring beam is about $0.2\times 0.2$~arcsec$^2$ for the MERLIN and about $0.015\times 0.015$~arcsec$^2$ for the EVN observations.
\subsection{MERLIN}
Only two of the 13 targets have not been detected. Both objects are IRAS sources, which suggests that the radio emission is extended and is likely dominated by extensive star formation. The other 11 sources are clearly detected (rms $\sim 0.07-0.4$~ mJy). In one source, we newly identified a nuclear double structure, while all other radio sources are unresolved on the MERLIN scale (i.e. $\sim 500$~pc at the average redshift of $z\sim 0.2$). On average 50\% of the FIRST flux is concentrated in the unresolved nuclear emission. The interpretation, however, is ambiguous, since for a few sources our new flux determination yields values that are higher than the previous measurement. Therefore, variability due to the nucleus or supernovae might play a role in these sources (see below).
\subsection{EVN}
Three sources have also been studied with western EVN and were clearly detected. They are also unresolved. The EVN scale corresponds to about 40~pc. Virtually 100\% of the corresponding MERLIN flux is contained in the EVN measurements.

\begin{figure}[t!]
\begin{center}
\includegraphics[width=12cm]{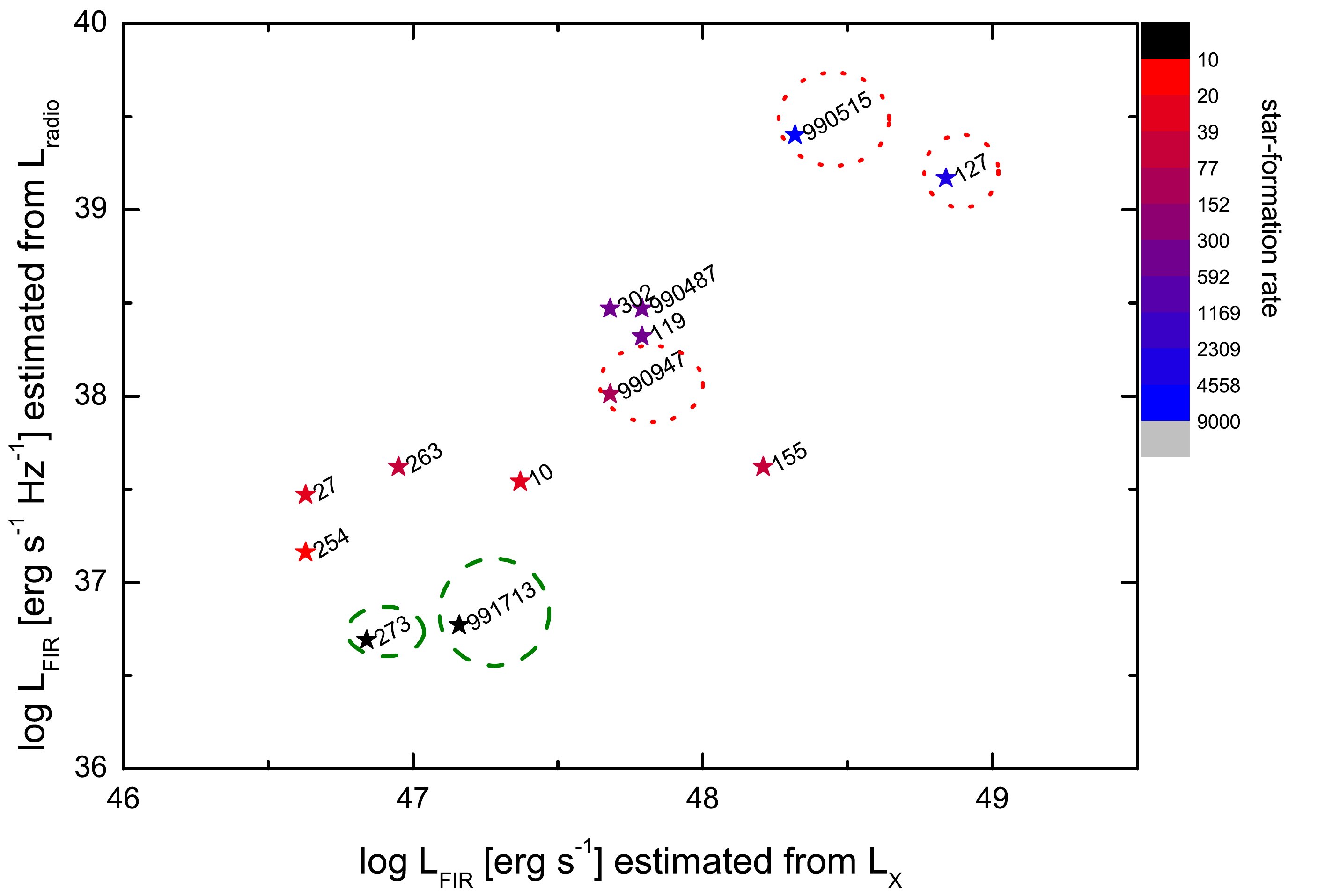}
\end{center}
\caption{Estimates of far-infrared luminosities and star-formation rates in order to clarify the nature of the radio emission. The dashed, green circles indicate the two IRAS detected sources that were fully resolved by our MERLIN observations.  Again, the red, dotted circles indicate targets that lie within galaxy clusters. See text for details. }
\label{fig:sfrate}
\end{figure}

To further investigate the nature of the radio emission, we test the hypothesis, whether the radio luminosity can be fully accounted for by star formation. For this purpose we derived star-formation rates using the 18~cm radio flux as a tracer of supernovae and therefore of star formation \cite{1992ApJ...388...82D}. The symbol color in Fig. \ref{fig:sfrate} corresponds to the estimated star formation rates, which are unrealistic high for many targets, in particular considering the nuclear apertures from which these values have been deduced from. Furthermore, we can estimate far-infrared (FIR) luminosities, another measure of ongoing star formation, from well known X-ray/FIR and radio/FIR correlations of normal or star-forming galaxies \cite{1992ApJ...388...82D,1992ARA&A..30..575C,2003ApJ...599..971H}.
The two measured IRAS luminosities are 2-3 magnitudes lower than the estimates, which indicates for a non-stellar nature of the radio and X-ray emission. Comparison with typical FIR luminosities from in-active galaxies \cite{2005ApJ...631..707P} shows that for given X-ray luminosity all targets are FIR under luminous. The same is true for the radio luminosity. We can also use the radio fluxes to compute the corresponding brightness temperatures. For the MERLIN observations, $<T_B>\sim 2\times 10^5$~K and for the EVN observations, $<T_B>\sim 5\times 10^7$~K. At least for the EVN data the values are too high to be dominated by star formation. Finally, the star-formation rates can be converted into supernova rates \cite{2001AJ....121..128H} and amount to about 15 supernovae per year. Again, for the EVN sources, these rates are somewhat too high considering the small volume (40~pc) in which they occur.

\section{Summary}
We have presented initial results from a radio-interferometric study of X-ray bright AGN. The detection rate is high and most of the sources are unresolved on the MERLIN (500~pc) scale. The three EVN observations are unresolved on the 40~pc scale and contain all the flux of the MERLIN measurements. Our observations show that the radio emission in these kind of sources is compact and likely related to the nuclear accretion processes, even considering the various spectral classes of objects and the connection of part of the X-rays to hot cluster gas in three cases. 

Future observations at other radio frequencies can provide further spectral information that can be incorporated in models of the nuclear emission processes \citep[e.g.,][]{blundell_origin_2007}. AO-assisted near-infrared follow-up studies, furthermore, have the power to complement the radio studies on scales comparable to that of MERLIN and can shed light on possible small-scale jet/interstellar medium interaction in these sources \citep[e.g.,][and references therein]{2007A&A...466..451Z}.

\ack This work is partly funded by the Deutsche Forschungsgemeinschaft via grant SFB 494. This work has benefited from research funding from the European Community's sixth Framework Programme under RadioNet R113CT 2003 5058187. We are grateful to Tom Muxlow and Giuseppe Cimo for their support on the data reduction.
\bibliography{zuther_j}

\providecommand{\newblock}{}
\begin{thebibliography}{10}
\expandafter\ifx\csname url\endcsname\relax
  \def\url#1{{\tt #1}}\fi
\expandafter\ifx\csname urlprefix\endcsname\relax\def\urlprefix{URL }\fi
\providecommand{\eprint}[2][]{\url{#2}}

\bibitem{page_submillimeter_2001}
Page M~J, Stevens J~A, Mittaz J~P~D and Carrera F~J 2001 {\em Science\/} {\bf
  294} 2516--2518

\bibitem{ho_search_1997}
Ho L~C, Filippenko A~V and Sargent W~L~W 1997 {\em \apj\/} {\bf 487} 568

\bibitem{2004ASPC..311...37W}
{Wilkes} B 2004 {\em AGN Physics with the Sloan Digital Sky Survey\/} ({\em
  Astronomical Society of the Pacific Conference Series\/} vol 311) ed
  {Richards} G~T and {Hall} P~B p~37

\bibitem{1978Natur.272..706S}
{Shields} G~A 1978 {\em \nat\/} {\bf 272} 706--708

\bibitem{2005ApJ...631..707P}
{Panessa} F, {Wolter} A, {Pellegrini} S, {Fruscione} A, {Bassani} L, {Della
  Ceca} R, {Palumbo} G~G~C and {Trinchieri} G 2005 {\em \apj\/} {\bf 631}
  707--719

\bibitem{1984ARA&A..22..319B}
{Bridle} A~H and {Perley} R~A 1984 {\em \araa\/} {\bf 22} 319--358

\bibitem{1992ARA&A..30..575C}
{Condon} J~J 1992 {\em \araa\/} {\bf 30} 575--611

\bibitem{2007A&A...467..519P}
{Panessa} F, {Barcons} X, {Bassani} L, {Cappi} M, {Carrera} F~J, {Ho} L~C and
  {Pellegrini} S 2007 {\em \aap\/} {\bf 467} 519--527

\bibitem{blundell_origin_2007}
Blundell K~M and Kuncic Z 2007 {\em \apj\/} {\bf 668} L103--L106

\bibitem{zuther08}
{Zuther} J {et al} 2008 {\em in prep.\/}

\bibitem{2007ApJS..172..634A}
{Adelman-McCarthy} J~K and {the SDSS consortium} 2007 {\em \apjs\/} {\bf 172}
  634--644

\bibitem{1999A&A...349..389V}
{Voges} W, {Aschenbach} B, {Boller} T, {Br{\"a}uninger} H, {Briel} U, {Burkert}
  W, {Dennerl} K, {Englhauser} J, {Gruber} R, {Haberl} F, {Hartner} G,
  {Hasinger} G, {K{\"u}rster} M, {Pfeffermann} E, {Pietsch} W, {Predehl} P,
  {Rosso} C, {Schmitt} J~H~M~M, {Tr{\"u}mper} J and {Zimmermann} H~U 1999 {\em
  \aap\/} {\bf 349} 389--405

\bibitem{2004ASPC..311..325Z}
{Zuther} J, {Eckart} A, {Straubmeier} C and {Voges} W 2004 {\em AGN Physics
  with the Sloan Digital Sky Survey\/} ({\em Astronomical Society of the
  Pacific Conference Series\/} vol 311) ed {Richards} G~T and {Hall} P~B p 325

\bibitem{2005sao..conf..375Z}
{Zuther} J, {Eckart} A, {Voges} W, {Bertram} T and {Straubmeier} C 2005 {\em
  Science with Adaptive Optics\/} ed {Brandner} W and {Kasper} M~E p 375

\bibitem{diss07}
{Zuther} J 2007 {\em Dissecting the host galaxies of Active Galactic Nuclei at
  high angular resolution\/} Ph.D. thesis University of Cologne

\bibitem{1992ApJ...388...82D}
{David} L~P, {Jones} C and {Forman} W 1992 {\em \apj\/} {\bf 388} 82--92

\bibitem{2003ApJ...599..971H}
{Hopkins} A~M, {Miller} C~J, {Nichol} R~C, {Connolly} A~J, {Bernardi} M,
  {G{\'o}mez} P~L, {Goto} T, {Tremonti} C~A, {Brinkmann} J, {Ivezi{\'c}} {\v Z}
  and {Lamb} D~Q 2003 {\em \apj\/} {\bf 599} 971--991

\bibitem{2001AJ....121..128H}
{Hill} T~L, {Heisler} C~A, {Norris} R~P, {Reynolds} J~E and {Hunstead} R~W 2001
  {\em \aj\/} {\bf 121} 128--139

\bibitem{2007A&A...466..451Z}
{Zuther} J, {Iserlohe} C, {Pott} J~U, {Bertram} T, {Fischer} S, {Voges} W,
  {Hasinger} G and {Eckart} A 2007 {\em \aap\/} {\bf 466} 451--466

\end{thebibliography}

\end{document}